\documentclass[12pt,preprint]{aastex}

\usepackage{natbib}
\usepackage{bm}
\usepackage{enumerate}
\usepackage{hyperref}
\usepackage{float}
\usepackage{amsmath}
\usepackage{amssymb}
\usepackage[latin1]{inputenc}
\newcommand{\brac}[1]{\langle #1 \rangle}

\newcommand{\mean}[1]{\overline{#1}}

\def\Ro{\mbox{\rm Ro}}

\def\Rs{R_{\odot}}

\usepackage{graphicx}
\usepackage{graphics}
\graphicspath{{fig/}{png/}}

\slugcomment{}

\shorttitle{Tachocline in TP balance}
\shortauthors{Guerrero et al.}
\begin{document}

\title{Understanding solar torsional oscillations from global dynamo models}

\author{G. Guerrero}
\affil{Physics Department, Universidade Federal de Minas Gerais,
Av. Antonio Carlos, 6627, Belo Horizonte, MG, Brazil, 31270-901}
\email{guerrero@fisica.ufmg.br}

\author{P.~K. Smolarkiewicz}
\affil{European Centre for Medium-Range Weather Forecasts, Reading RG2 9AX, UK}
\email{smolar@ecmwf.int}

\author{E. M. de Gouveia Dal Pino}
\affil{Astronomy Department, Universidade de S\~{a}o Paulo, IAG-USP, 
Rua do Matão, 1226, São Paulo, SP, Brasil, 05508-090}
\email{dalpino@astro.iag.usp.br}

\author{A.~G. Kosovichev}
\affil{New Jersey Institute of technology, Newark, NJ 07103,USA}
\email{alexander.g.kosovichev@njit.edu}

\and
\author{N. N. Mansour}
\affil{NASA, Ames Research Center, Moffett Field, Mountain View, CA 94040, USA}
\email{Nagi.N.Mansour@nasa.gov}

\newpage

\begin{abstract}
The phenomenon of solar ``torsional oscillations" (TO) represents
migratory zonal flows associated with the solar cycle. These flows are 
observed on the solar surface and, according to helioseismology, extend
through the convection zone.  We study the origin of the TO using 
results from a global MHD simulation of the solar interior that reproduces
several of the observed 
characteristics of the mean-flows and magnetic fields.
Our results indicate that the magnetic tension (MT) in the tachocline 
region is a key factor for the periodic changes
in the angular momentum transport that causes the TO. The torque
induced by the MT at the base of the convection zone
is positive at the poles and negative at the equator. A 
rising MT torque at 
higher latitudes causes the poles to speed-up, whereas a declining
negative MT torque at the lower latitudes causes the 
equator to slow-down. These changes in the zonal flows propagate
through the convection zone up to the surface.   Additionally, our results suggest 
that it is the magnetic field at the tachocline that modulates
the amplitude of the surface meridional flow rather than the opposite
as assumed by flux-transport dynamo models of the solar cycle. 
\end{abstract}


\section{Introduction}

The Sun exhibits a periodic variation of its angular velocity that
is about $\pm 0.5$\% of the average rotation profile. These so-called 
torsional oscillations (TO) represent
acceleration and deceleration of the zonal component of the
plasma flow during the solar cycle \citep{HL80,KS97,T+00,H+00,AB01,H+05,V+02}.
Two branches of the oscillations have been observed. The first branch, migrating 
from ~40$^o$ latitude toward the equator, appears
a few years prior to the start of a solar cycle and disappears by the
end of the cycle. The second branch, at higher latitudes, migrates
toward the polar regions. For the solar cycles 22 and 23, the amplitude
of the polar branch was larger than that of the equatorial branch.
For  cycle 24, the polar branch is significantly weaker than it
was in the previous cycles 
\citep{ZKB14,KHG14,KZ16}. Additionally, recent helioseismology observations
showed that the TO correlate with the variations of the meridional
flows \citep{ZKB14,KGH15}. 
The evident dependence of the TO on the solar activity cycle
provides a unique opportunity to explore the interaction between
large-scale magnetic field and flows. Furthermore, understanding
the nature of TO could provide a way to infer the distribution
of magnetic fields below the solar photosphere.

The origin of the TO is still unclear. It is
puzzling that the equatorward branch starts before the
beginning of the magnetic cycle. The magnetic feedback via the Lorentz
force on the plasma flow is one possible 
explanation motivated by mean-field turbulent dynamo models 
\citep{Yo81,KR81,CTMT00,CMT04}.  
In a flux-transport dynamo model, where the source of the poloidal field
is non-local and depends on the buoyancy of magnetic flux tubes, 
\cite{R07} explained the high latitude branch of the oscillations as a
result of the magnetic forcing, while arguing that the equatorial branch 
has a thermal origin. Indeed, \cite{Sp03} considered the proposition 
that oscillations are driven by temperature variations at
the surface, which are due to the enhanced emission of small magnetic structures.
In contrast, recent numerical simulations by \cite{BCRS12} showed that 
the TO can be driven via the magnetic modulation of the angular momentum 
transport by the large-scale meridional flow.

In this work, we study the origin of TO using a similar numerical 
nonlinear global dynamo simulation that
captures several important characteristics of the solar cycle 
\citep[][hereafter Paper~I]{GSDKM16a},  and qualitatively reproduces the 
surface pattern of angular velocity variations. The results from Paper 
I that are relevant to this work are summarized in \S\ref{s.p1}; the new 
analysis and results are presented in \S\ref{s.r}; and finally, 
we conclude in \S\ref{s.c}. 

\section{Solar global dynamo model}
\label{s.p1}

In Paper~I, we have performed global MHD simulations of spherical turbulent
rotating convection and dynamo using the EULAG-MHD code \citep{SC13,GSKM13b}, 
a spin-off of the hydrodynamical code EULAG predominantly used in
atmospheric and climate research \citep{PSW08}. The goal
of Paper~I was to compare dynamo models that consider 
only the convection zone (models CZ) with models that also include 
a radiative zone (models RC) and, thus, naturally develop a tachocline 
at the interface between the two layers. In particular, the simulation 
case RC02 rotating with the solar angular velocity,  
described in Paper~I, results in a pattern of the
differential rotation comparable with the solar observations 
(Fig. \ref{fig.to}(a)). The meridional circulation
exhibits two or more cells in the radial direction at lower latitudes.
In a thin uppermost layer,  the model shows a poleward
flow for latitudes above $\sim30^o$ latitude and an equatorward flow
for latitudes $\lesssim 30^o$ (Fig. \ref{fig.to}(a)). 
This pattern is the result of a negative axial torque due to the 
Reynolds stresses, which sustains the near-surface shear layer (NSSL) 
while driving meridional flows away from the rotation axis, 
and also of the impermeable boundary condition which enforces a counterclockwise 
(clockwise) circulation in the northern (southern) hemisphere.
This poleward flow does not appear at equatorial
regions because of the large convective ``banana-shaped''  cells.
In the model, the dynamo 
mechanism results in oscillatory magnetic fields that are indicated 
by white contour lines in both panels of Fig. \ref{fig.to}(b). 
The magnetic field in the convection zone evolves
in a highly diffusive regime with 
$\eta_{\rm t} \sim 10^9 {\rm m}^2{\rm s}^{-1}$ (in agreement with 
mixing-length theory). In spite of this large value, the dynamo
full cycle period is $\sim 34$~year. The cycle period of a similar dynamo
model without tachocline (model CZ02 in Paper~I) has a cycle period of
$\sim 2$~year. 
The reason for this difference is that, unlike model CZ02, where the
turbulent diffusivity in the convection zone sets the period, 
in model RC02 the evolution of the large-scale magnetic field is 
governed by the magnetic field seated in the tachocline and the 
radiative zone. Magneto-shear instabilities occurring in this region 
develop non-axisymmetric modes that periodically exchange energy with 
the large-scale magnetic field. The time-scale at which this exchange 
occurs determines the dynamo cycle period (see the sections 3.3 and 
3.4 and  panel (b) of Figure 9 in Paper I).

\section{Understanding torsional oscillations}
\label{s.r}

\begin{figure}[H]
\begin{center}
\includegraphics[width=0.40\columnwidth]{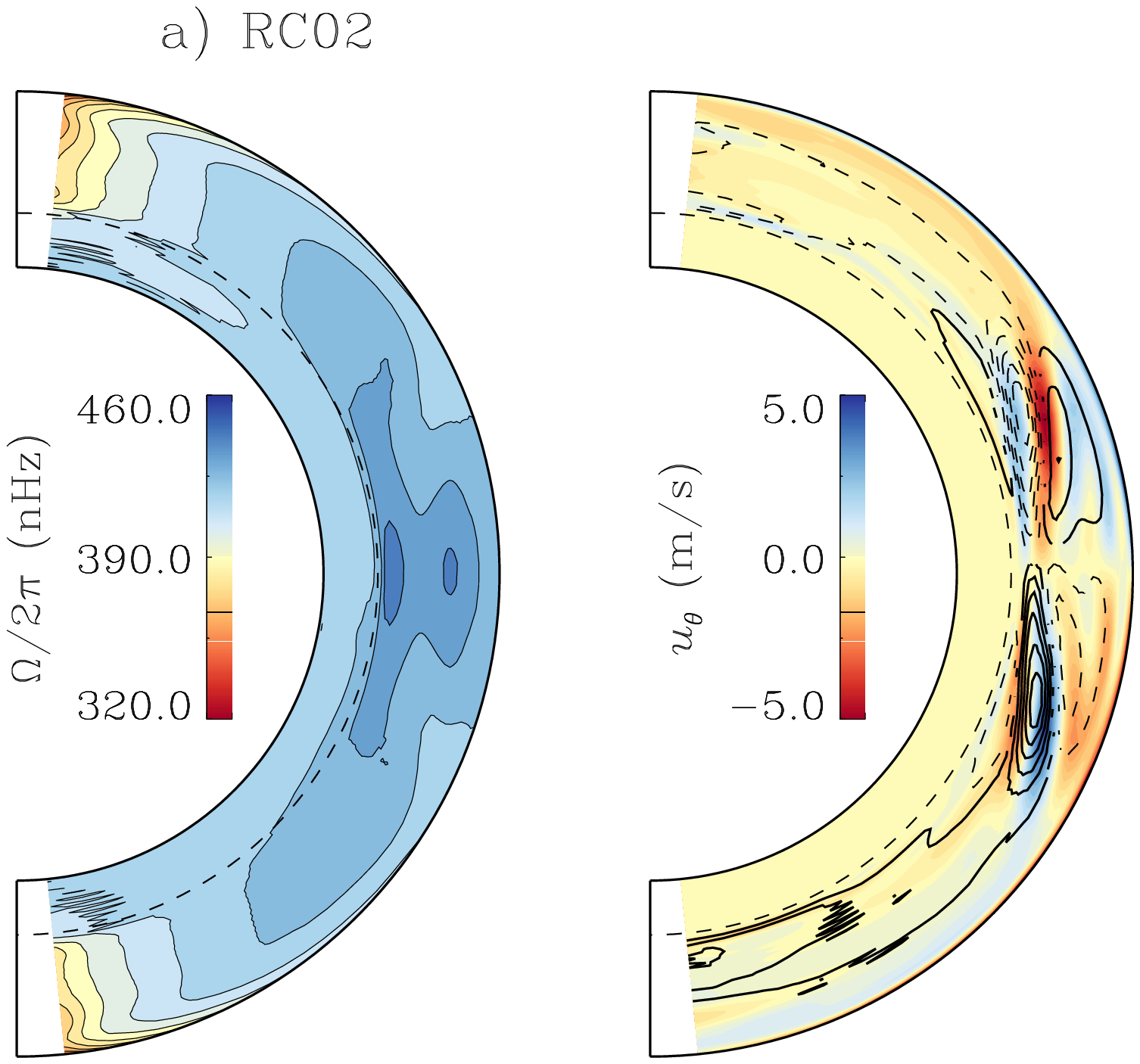}
\includegraphics[width=0.54\columnwidth]{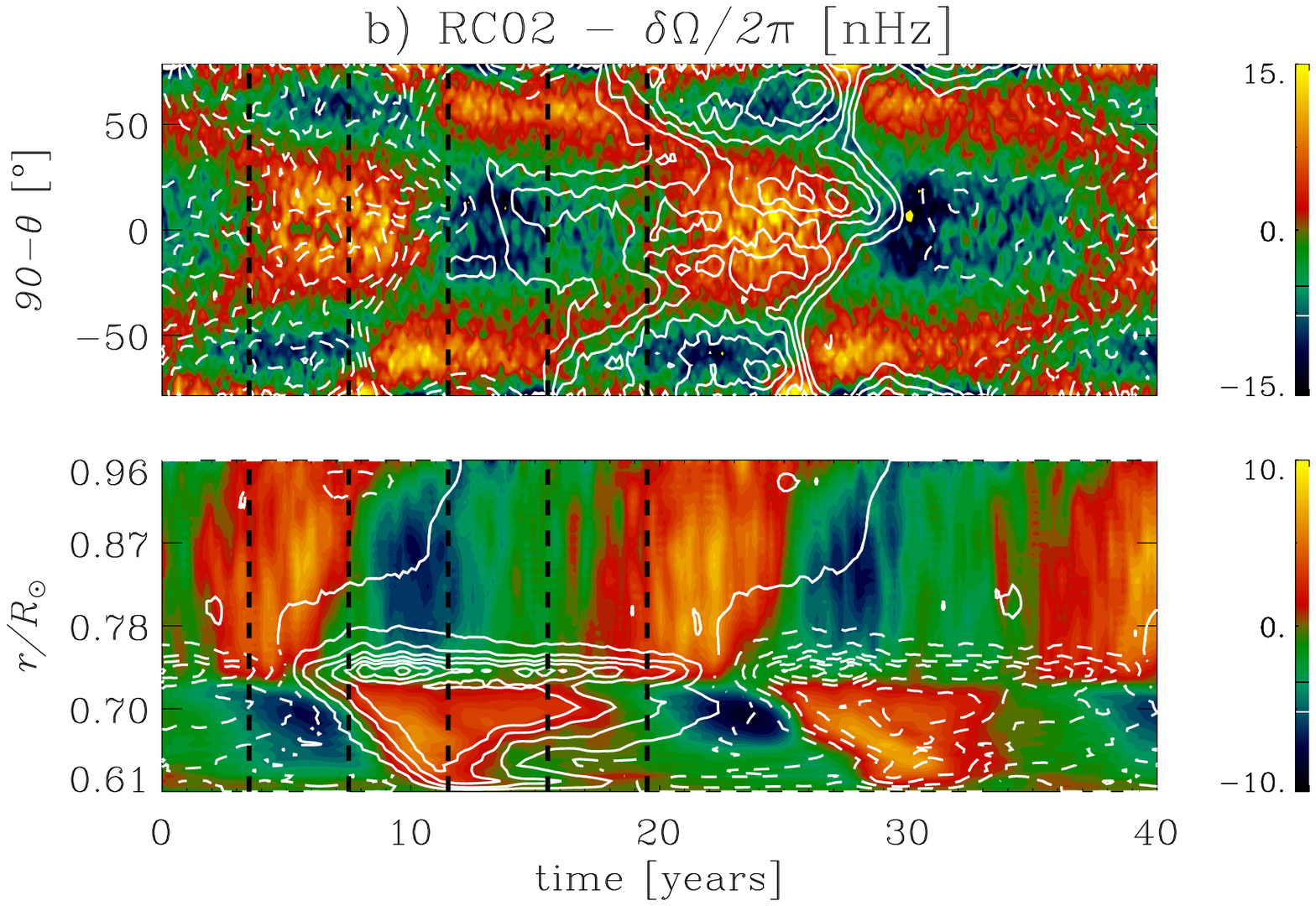}
\caption{Left: (a) meridional distribution of the angular velocity 
and the meridional circulation for the model RC02. The profiles correspond
to zonal and temporal averages over $\sim 10$ year during the steady 
phase of the simulation. Right: (b) time-latitude diagram for 
$r=0.95\Rs$ (upper panel), 
and time-radius diagram for
$30^o$ latitude (bottom panel) of 
$\delta \Omega (r,\theta,t)/2\pi = (2\pi \varpi)^{-1} 
(u_{\phi}(r,\theta,t)-\mean{u}_{\phi}(r,\theta))$. The 
continuous (dashed) white line 
contours depict the positive (negative) toroidal magnetic field 
strength also shown in Fig. 6b of Paper~I.}
\label{fig.to}
\end{center}
\end{figure}

The oscillatory dynamo model RC02 described above has a natural
evolution of the angular velocity with faster and slower regions migrating
as the magnetic cycle progresses, as 
shown in Fig. \ref{fig.to}(b). These panels depict 
$\delta \Omega (r,\theta,t) = (2\pi \varpi)^{-1} (u_{\phi}(r,\theta,t)-
\mean{u}_{\phi}(r,\theta))$, where $\varpi=r\sin\theta$ and 
$(2\pi \varpi)^{-1} \mean{u}_{\phi}$ is the zonal and temporal average 
shown in Fig. \ref{fig.to}(a). In the Figure \ref{fig.to}(b) the upper panel
shows the oscillations in time and latitude at $r=0.95\Rs$. The bottom 
panel shows the time-radius evolution of $\delta\Omega$ at $30^o$ latitude.  
Red (green) filled  contours represent speed-up (slow-down) 
of the angular velocity.   We notice, first, that at 
surface levels ($r=0.95\Rs$) the model has a pattern with polar and 
equatorial branches that resemble quite well the solar observations
\citep[cf. Fig. 25 of ][]{Howe09}. In the radial
direction, the oscillations appear in the entire convection zone and the 
stable layer. The amplitude of the oscillations is up to 5\% of the 
angular velocity, i.e., a few times larger than the observed ones. 
In a turbulent mean-field model, where the magnetic feedback is mediated 
solely by the Lorentz force, \cite{CMT04} found that the amplitude 
of the oscillations depends on the amplitude of the $\alpha$-effect.
It is possible to verify this finding in global simulations 
by varying the Rossby number.  This changes the kinetic
helicity of the simulations (the $\alpha$-effect) but also  
modifies the shear profile (i.e., the $\Omega$-effect) and occasionally 
results in different dynamo modes (see Paper I). We have performed three
complementary simulations where we changed the polytropic index of
the convection zone such that 
the Rossby numbers of the simulations are slightly smaller/larger than 
for the model RC02 ($\Ro=0.069$) but still resulting in oscillatory 
dynamos. All of the cases exhibit a pattern of TO similar to that in 
Fig.~\ref{fig.to}(b). The results indicate that the amplitude of the 
TO in the equatorial region has a local maximum for $\Ro = 0.070$.
For the polar regions, the amplitude of the TO is smaller for the smaller 
values of $\Ro$, but remains approximately constant for the larger values.

In Fig. \ref{fig.to}(b) a speed-up of the zonal flow at higher 
latitudes during the minimum of the toroidal field can be observed. This branch 
propagates toward the equator. When the magnetic field is strong at higher 
latitudes, there is a branch of less rapidly rotating plasma that also 
propagates toward the equator.
At lower latitudes,  the change between speed-up and
slow-down occurs roughly when the toroidal field reverses polarity. 
In the tachocline, the field is stronger at the starting phase of the magnetic 
cycle.  During this phase the zonal flow is accelerated (bottom panel of Fig. 
\ref{fig.to}(b)). It decelerates during the declining and reversal stages of 
the cycle. 

To look for a physical explanation for these features, we start by analyzing
the angular momentum balance resulting from the different transport 
mechanisms. After multiplying the zonal component of the momentum equation 
(Eq. 3 in Paper 1) by $\varpi$, and then making a 
mean-field decomposition,  we obtain the equation for the angular 
momentum evolution:
\begin{equation}
 \frac{\partial ({\rho}_s \mean{u}_{\phi})}{\partial t}
     =-\frac{1}{\varpi}                 
      \nabla\cdot \left( \varpi \left[ 
      \rho_s (\mean{u}_{\phi}+\varpi \Omega_0)\mean{\bm u}_{\rm m} + 
      \rho_s \mean{u_{\phi}' {\bm u}'_{\rm m}} 
      -\frac{1}{\mu_0} \mean{B}_{\phi}\mean{{\bm B}}_{\rm p}
      -\frac{1}{\mu_0} \mean{b_{\phi}'{\bm b}_{\rm p}'} \right]  \right ),
      \label{eq.amb} 
\end{equation}
where, $\mean{\bm u}_{\rm m}$, ${\bm u}'_{\rm m}$, $\mean{\bm B}_{\rm p}$ and
${\bm b}'_{\rm p}$ denote the mean and turbulent meridional ($r$ and $\theta$) 
components of the velocity and magnetic fields, respectively. Because the 
lhs of Eq. \ref{eq.amb} is not in equilibrium but oscillates in time, the 
eight terms under the divergence in the rhs - 
\begin{eqnarray}                                                                                            
 \label{eq.amf1}
 {\cal F}_r^{\rm MC}&=&\rho_s \varpi (\mean{u}_{\phi}+\varpi \Omega_0)   \mean{u}_r, \\\nonumber
 {\cal F}_{\theta}^{\rm MC}&=&\rho_s \varpi (\mean{u}_{\phi}+\varpi \Omega_0) \mean{u}_{\theta},\\\nonumber
 {\cal F}_{r}^{\rm RS}&=&\rho_s \varpi \mean{{u}_{\phi}' {u}_{r}'}, \\\nonumber
 {\cal F}_{\theta}^{\rm RS}&=&\rho_s \varpi \mean{{u}_{\phi}' {u}'_{\theta}}, \\\nonumber
 {\cal F}_r^{\rm MT}&=&-\frac{\varpi}{\mu_0} \mean{B}_{\phi} \mean{B}_r, \\\nonumber
 {\cal F}_{\theta}^{\rm MT}&=&-\frac{\varpi}{\mu_0} \mean{B}_{\phi} \mean{B}_{\theta},\\\nonumber
 {\cal F}_{r}^{\rm MS}&=&-\frac{\varpi}{\mu_0} \mean{{b}_{\phi}' {b}_{r}'}, \\\nonumber
 {\cal F}_{\theta}^{\rm MS}&=&- \frac{\varpi}{\mu_0} \mean{{b}_{\phi}' {b}'_{\theta}},
\end{eqnarray}       
namely, the fluxes of angular momentum due to meridional circulation (MC), 
Reynold stresses (RS), the axisymmetric (or magnetic tension MT) and turbulent 
Maxwell stresses (MS) - should account for such variations.  The net angular momentum transport 
is then computed \citep[see][]{BMT04,BCRS12} as
\begin{eqnarray}
 I_r(r)=\int_0^{\pi} {\cal F}_r(r,\theta) r^2 \sin\theta d\theta \\\nonumber
 I_{\theta}(\theta)=\int_{r_b}^{r_t} {\cal F}_{\theta}(r,\theta) r \sin\theta dr,
\end{eqnarray}
where $r_b = 0.61 \Rs$, $r_t = 0.96 \Rs$ and:
\begin{eqnarray}
 \label{eq.amf2}
 {\cal F}_r={\cal F}_r^{\rm MC}+{\cal F}_{r}^{\rm RS}+{\cal F}_r^{\rm MT}+{\cal F}_{r}^{\rm MS} \\\nonumber
 {\cal F}_{\theta}={\cal F}_{\theta}^{\rm MC}+{\cal F}_{\theta}^{\rm RS}+{\cal F}_{\theta}^{\rm MT}+{\cal F}_{\theta}^{\rm MS}.
\end{eqnarray}

\begin{figure}[H]
\begin{center}
\includegraphics[width=0.9\columnwidth]{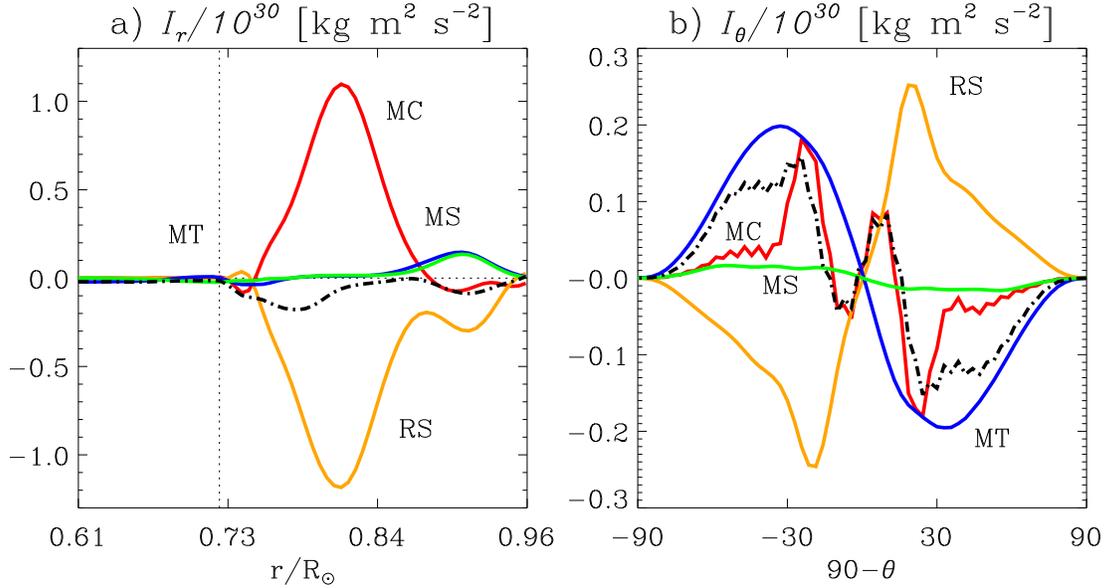}
\caption{Integrated angular momentum fluxes for model RC02 as defined by Eqs. 
(\ref{eq.amf1}) to (\ref{eq.amf2}). 
The red, orange, blue, and green lines correspond to the MC 
(meridional circulation), RS (Reynolds stress), 
MT (magnetic tension),  and MS (Maxwell stresses) fluxes, respectively.
The black dot-dashed lines are the total contribution in each direction.}  
\label{fig.amb}
\end{center}
\end{figure}

Fig. \ref{fig.amb} shows the integrated radial (left) and latitudinal 
(right) profiles of the angular momentum fluxes of the rhs of Eq. \ref{eq.amf2}.
Red, orange, blue, and green lines correspond to the MC, RS, MT, and MS fluxes, 
respectively. Since the lhs of Eq. (\ref{eq.amb}) is different from zero
during the magnetic cycle, in order to investigate a steady state balance we average the 
integrated fluxes for two complete magnetic cycles. The black dotted-dashed lines 
correspond to the sum of all of the integrated magnetic fluxes in each direction. 
By interpreting the information in this
figure, it is possible to disentangle the different contributions 
to the transport of angular momentum.

The radial transport shows a larger contribution from the Reynolds stress 
component, which is balanced by the MC term. 
It is evident from the black dotted-dashed line in Fig. \ref{fig.amb}(a)
that the sum of the integrated fluxes  balances well in the convection zone. 
The upper limit for the viscous contribution to the angular momentum transport, 
i.e., the residual of ${\cal F}_r$ (see Eq. \ref{eq.amf2}), is less than 
10\% of the values of the RS or MC fluxes in this region.

Both magnetic contributions seem to be negligible in most of the domain
except for the interface between the stable and the unstable layers and for
the near-surface shear layer. In the stable region, the viscous 
angular momentum flux is significant and balances the magnetic fluxes.  
The amplitude of these components in the radiative zone is less 
than 2\% of the value in the convection zone. However, it is sufficient 
for transporting angular momentum into the stable region, ultimately making 
this layer to rotate faster on average than the reference frame
(Fig. \ref{fig.to}(a)).  Because periodic changes in the angular
momentum transport in the radiative zone are solely due to the magnetic 
fluxes, it is clear that the TO in the radiative zone must have magnetic origin. 

Figure \ref{fig.amb}(b) helps to explain the solar-like 
differential rotation pattern observed in Fig. 
\ref{fig.to}(a). First of all, the dominant term is the Reynold stress flux,
which is positive (negative) in the northern (southern) hemisphere. This 
means that the angular momentum transport is equatorward. The dominant RS is 
a robust feature of global models with solar-like differential rotation
\citep[e.g.][]{GSKM13b,FM15}. Opposed to this transport are the meridional
circulation, the magnetic tension and the Maxwell stress fluxes. 
The amplitude of the latitudinal integrated fluxes (Fig. \ref{fig.amb}(b))  
is about 30\% smaller than the radial fluxes, and the viscous contribution 
(which is of the same order as that in Fig. \ref{fig.amb}(a)) appears to be 
important for the latitudinal transport. Because at lower latitudes the RS 
and MT fluxes balance each other, the viscous flux clearly compensates the 
variation of the meridional circulation. At higher latitudes, the 
sum of the MC and MT fluxes should balance the RS flux. However, 
due to temporal variations of both quantities, they are not in balance. 
The sub-grid scale (SGS) viscous flux balances this difference.
\footnote{This is in fact how the SGS viscosity, in this case implicit, works; 
it is larger where the local derivatives of the velocity field
are important \citep{MRG06,PSMW09}.} As shown later in 
Fig. \ref{fig.tstor}, the variance of the RS flux over the cycle is minimal.
This suggests that the TO are of magnetic origin.
For instance, they may be driven directly through the 
large-scale magnetic torque, but also indirectly via the transport of 
angular momentum by meridional flow modulated by the magnetic cycle, as 
suggested by \cite{BCRS12}.

\begin{figure}[H]
\begin{center}
\includegraphics[width=0.99\columnwidth]{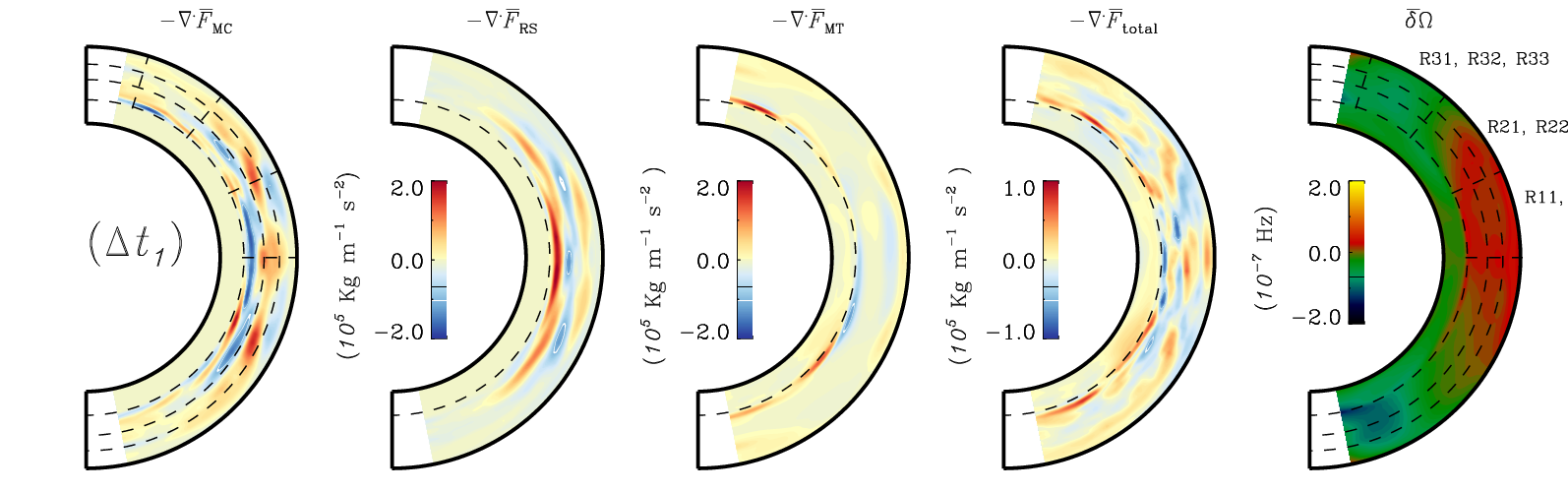}\\
\includegraphics[width=0.99\columnwidth]{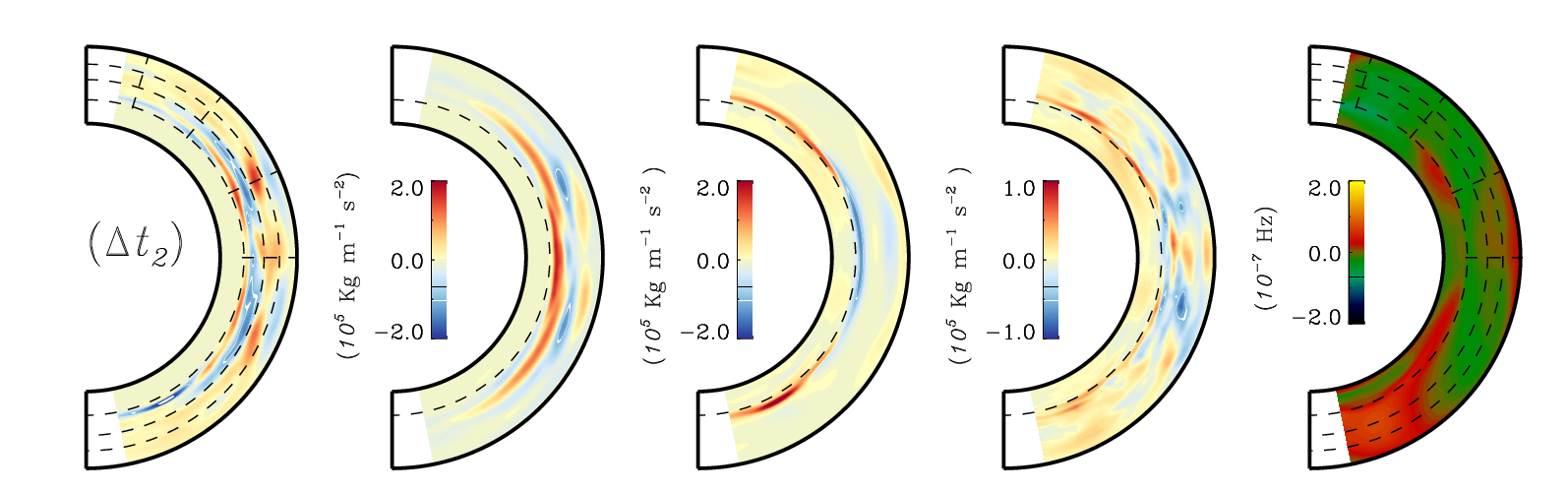}\\
\includegraphics[width=0.99\columnwidth]{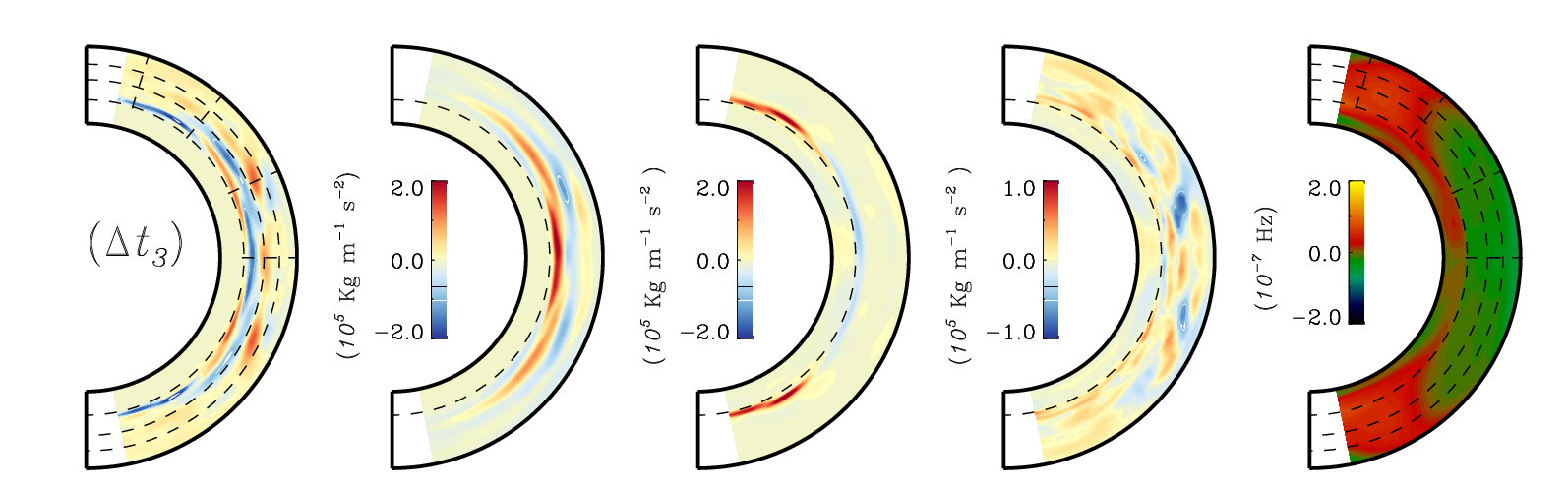}\\
\includegraphics[width=0.99\columnwidth]{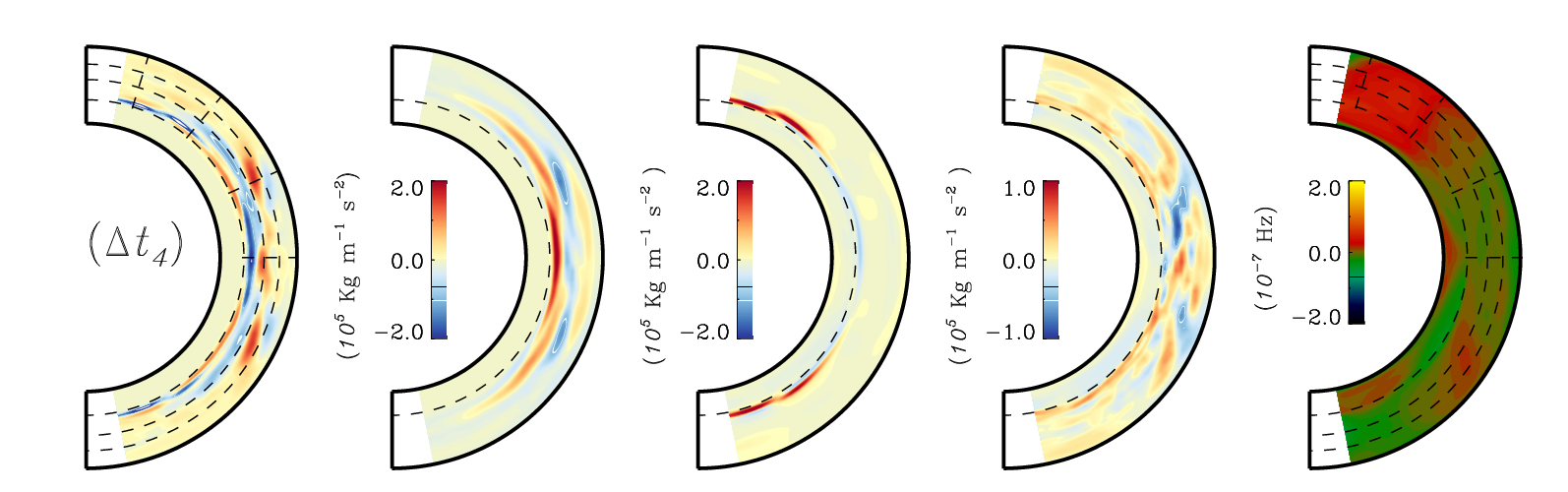}\\
\caption{Meridional profiles of the total axial torques and 
$\mean{\delta \Omega}$ in four different time intervals
corresponding to the different phases of the magnetic cycle
depicted with black dashed lines in Fig.~\ref{fig.to}(b). 
For a better comparison with the available observational 
results, in the figure we present latitudes only up to $85^o$.
}
\label{fig.Fth}
\end{center}
\end{figure}

\begin{figure}[H]
\begin{center}
\includegraphics[width=0.99\columnwidth]{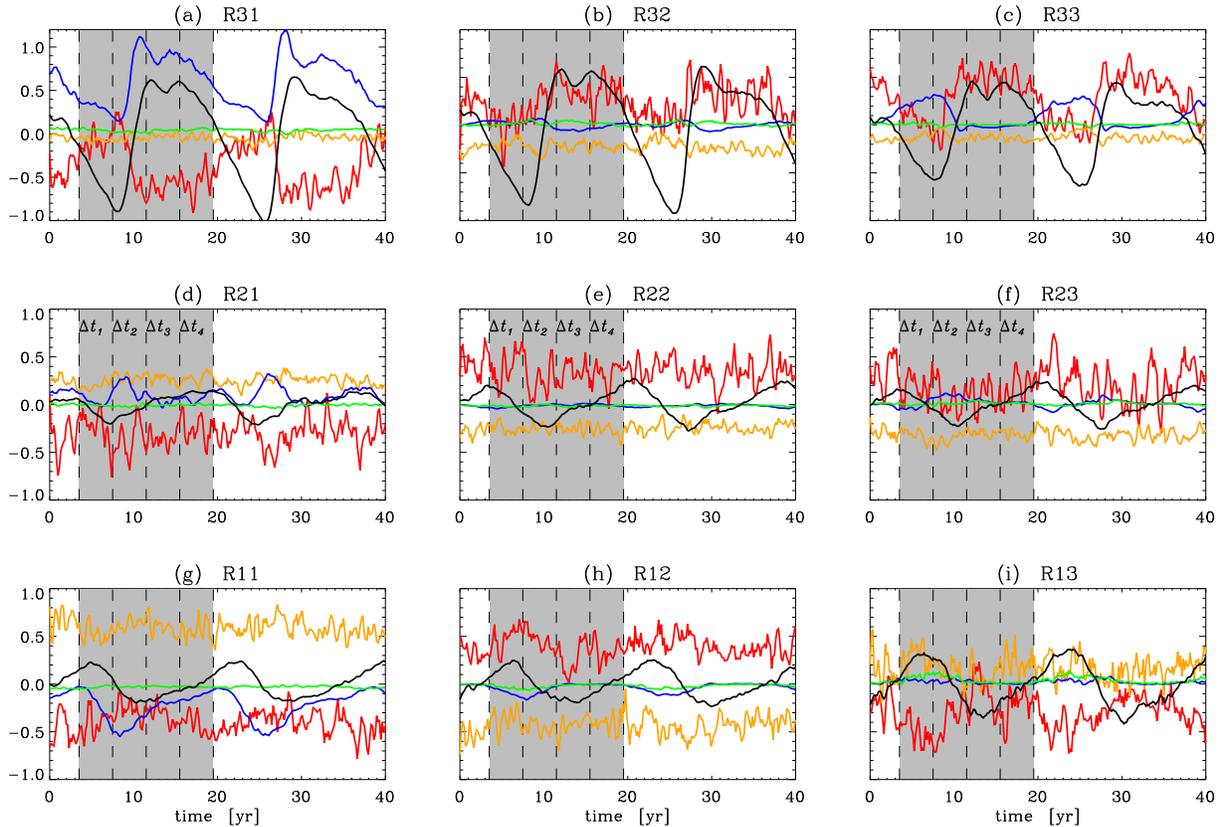}
\caption{Time evolution of the axial torques computed for different latitudinal
and radial regions labeled in the upper row of Fig. \ref{fig.Fth}.
Red, orange, blue and green lines correspond
to the MC, RS, MT and MS axial torques normalized to the local maximum value
of $\brac{\nabla \cdot \cal{F}_{\rm MC}}$. The black line shows the evolution
of $\brac{\delta \Omega}$ normalized to $10^{-7}$. The angular brackets 
mean volume averages over each region.}
\label{fig.tstor}
\end{center}
\end{figure}

Next,
we compare the meridional profiles of the torsional
oscillations with those of the axial torques, 
$-\mean{\nabla \cdot \cal{F}}$, where $\cal{F}$ are the eight quantities 
of Eq. \ref{eq.amf1} (Fig~\ref{fig.Fth}).  The columns, 
from left to right, correspond to the MC, RS, and MT torques, 
the total axial torque, and  
$\mean{\delta \Omega}$ as defined above, respectively. 
The rows correspond to four different time averages (of 4 years each)
in four different phases of a full magnetic cycle. These
phases starting at $t=3.5$ year are depicted by black dashed lines in Fig. 
\ref{fig.to}(b). For the torques, the red and blue contours 
represent positive (entering into the page) and negative (out of the page) 
directions, respectively. We do not present $-\mean{\nabla \cdot \cal{F}_{\rm MS}}$,
because its contribution is negligible in most of the domain. 
For the TO (rightmost panel), like in Fig. \ref{fig.to}(b), the red 
(green) contour
levels indicate a speed-up (slow-down) of the rotation. We notice 
that the meridional profiles of TO in different cycle phases 
also closely resemble the observations 
\citep[see Fig.~8 of][for comparison]{H+05}.

As discussed in the literature  \citep[e.g.,][]{FM15}, 
the balance of these torques is responsible for the maintenance of the 
mean-flows. As expected from Fig. \ref{fig.amb},
$-\mean{\nabla \cdot \cal{F}_{\rm MC}}$ and
$-\mean{\nabla \cdot \cal{F}_{\rm RS}}$ tend to balance 
each other in the bulk of the convection zone. Whenever
the RS axial torque is negative (positive), it will induce a 
meridional flow away from (toward) the rotation axis. This
explains well the two main MC cells
at lower latitudes, the deeper cell being counterclockwise and
the shallow one being clockwise \citep[see e.g.,][for a 
complete analysis]{FM15}.  
Despite the fact that our simulation is MHD, the profiles 
of $-\mean{\nabla \cdot \cal{F}_{\rm MC}}$ and 
$-\mean{\nabla \cdot \cal{F}_{\rm RS}}$ compare well with those
obtained by, e.g., \cite{BMT11,FM15}. 
Nevertheless, the profile of $-\mean{\nabla \cdot \cal{F}_{\rm RS}}$ 
presented here has negative values near the surface and at latitudes
above $30^o$.  As mentioned before, this negative
torque is ultimately responsible for the formation of the NSSL 
\citep[see also][]{MH11,HR15}.  

The profile of $-\mean{\nabla \cdot \cal{F}_{\rm RS}}$ 
does not show significant changes during the magnetic cycle.
On the other hand, $-\mean{\nabla \cdot \cal{F}_{\rm MC}}$ 
exhibits regions of strong periodically varying torque at the base 
of the convection zone and variations of its amplitude at 
the surface. These changes are a response to the magnetic 
torques, mainly to $-\mean{\nabla \cdot \cal{F}_{\rm MT}}$,
which peaks at the rotational shear layers.  The latitudinal 
distribution of $-\mean{\nabla \cdot \cal{F}_{\rm MT}}$ at
the tachocline has 
positive values at higher latitudes and negative values 
at the equator. The change in amplitude of 
$-\mean{\nabla \cdot \cal{F}_{\rm MC}}$ and
$-\mean{\nabla \cdot \cal{F}_{\rm MT}}$ seems to be associated
with the TO presented in the rightmost column. 
This is reflected in the total axial torque
(fourth column of Fig. \ref{fig.Fth}) which in the first
two phases, $\Delta t_1$ and $\Delta t_2$ shows bluish
(negative torque) levels at higher latitudes, accounting for the
slowed-down poles. During the two last phases, $\Delta t_3$ 
and $\Delta t_4$, $-\mean{\nabla \cdot \cal{F}_{\rm total}}$
has reddish (positive torque) levels at higher latitudes
when the poles speed-up and the equator slows-down. 

To explore the locality and causality issue more deeply, we have divided 
the first meridional quadrant of the domain into nine regions, 
which can be observed in the first panel of Fig. \ref{fig.Fth}.
The equatorial regions, R11, R12, and R13, span from 
$0^o \le \theta < 25^o$ 
latitude and $0.71 \le r/\Rs < 0.81$, $0.81 \le r/\Rs < 0.88$, and 
$0.88 \le r/\Rs \le 0.96$ in radius, respectively; intermediate latitude 
regions, R21, R22, and R23, span from $25^o \le \theta < 50^o$;  and high 
latitude regions, R31, R32, and R33, span from 
$50^o \le \theta < 75^o$, all with the same radial extents. For each 
region we have computed the volume average of the axial torques but did
not average in time, in order to study the time evolution of the four 
quantities and assess their relative importance. 
In the results, presented in Fig. \ref{fig.tstor}, the time series
lines follow the same color pattern as in Fig. \ref{fig.amb},
i.e., the red, orange, blue, and green lines correspond to the MC, 
RS, MT, and MS axial torques, respectively. These quantities are normalized to the
maximum local value of $-\brac{\nabla \cdot {\cal F}_{\rm MC}}$
(the angular brackets indicate volume averages over each reagion).  
The black lines correspond to $\brac{\delta \Omega}$ normalized
to $10^{-7}$ Hz.  The gray shaded region indicates the four time
intervals presented in Fig. \ref{fig.Fth}. These time series explain 
the origin of the TO. 

In region R31 (pole-bottom, Fig. \ref{fig.tstor}(a)),
there is a clear correlation between the MT torque and $\brac{\delta \Omega}$.
At $\Delta t_1$ 
the positive MT torque is decreasing while the negative MC torque 
increases,  balancing the angular momentum. The amplitude 
of the MT torque is higher,  leading to a slow-down of the angular velocity. 
In $\Delta t_2$ the MT torque quickly increases and $\Omega$ speeds-up 
in spite of the decline of MC. The phases $\Delta t_3$ and $\Delta t_4$ 
correspond to the maximum (minimum) and to the decline (rise) of the MT (MC) 
torque, respectively, and are associated to the 
slow-down of $\Omega$.  The phase relation between the blue
and the black lines leaves no doubt that the MT causes
the TO. It is remarkable that, in this
latitude range, the TO pattern remains more or less the same in radius,
with only minor changes of the amplitude. 
In region R32 (pole-bulk, panel (b)) the RS torque is 
negative and MC is positive. The latter shows variations that seem to
follow the TO.  
In R33 (pole-top, panel (c)), the MC
torque is mostly positive and clearly follows the TO, becoming negative 
in its minima.  The MT torque in this zone is due to the 
magnetic field generated in the NSSL and anti-correlates with the TO which 
seems to conserve the imprints produced in it by the MT torque at the bottom
layers, in R31.

In the intermediate regions R21, R22, and R23 (Fig. \ref{fig.tstor}(d)-(f)), 
the torques exhibit less variation while $\brac{\delta\Omega}$ conserves 
roughly the same oscillation pattern but with some phase delay (of about $\pi/4$)
and smaller amplitudes than that 
acquired at higher latitudes. In R22 and R23, the changes
from positive to negative occur at the beginning of $\Delta t_2$, i.e.,
4 years after the transitions in R32 and R33. This is a consequence of the
migration of the positive (negative) MT toward the equator (poles) 
observed in Fig. \ref{fig.Fth}. This migration explains the propagation
of the TO.

Finally, in R11 (equator-bottom, panel (g)), the time series indicate 
a similar behavior to what is observed in 
the region R31, i.e., there is a clear correlation between the MT torque and
the TO. Besides, we note that in this case the MT and MC torques are 
negative and, evidently, oscillate in anti-phase. The sign change of the MT torque
from the poles to the equator reflects the fact that, in the dynamo,
the mean toroidal and poloidal fields have certain phase difference. 
Opposite to region R31, in R11 $\brac{\delta\Omega}$ is positive 
during $\Delta t_1$ (see also the two upper rows of Fig. \ref{fig.Fth}).
It also has lower amplitude. The evolution of the TO curve follows that
of the MT torque, i.e., it rises slowly 
and declines rapidly. The observed correlation suggests that this 
equatorial oscillation is driven by the MT.
It also propagates upwards,
conserving a similar shape in regions R12 and R13. 
In panel (i), region R13 (equator-top), the MC torque oscillates 
with a large amplitude which 
does not seem to be compensated by any other torque. This imbalance
(actually seen also in R33 and R23) is expected, as this is the 
fraction of the domain where the numerical viscous contribution to
the angular momentum transport is relevant (see Fig. \ref{fig.amb}a).
Once again, the MC oscillations seem to be connected to the TO and
unrelated to any other torque variation in the same region. 

Besides the changes with the periodicity of the 
magnetic cycle,  the MC torque also depicts erratic short term fluctuations. 
On the other hand, the evolution of $\brac{\delta \Omega}$ is less noisy, which 
is also the case for the magnetic torque $-\brac{\nabla \cdot {\cal F}_{\rm MT}}$. 
This supports the argument that the TO
are induced by the MT at the base of the convection
zone.  Thus, the observed variation of the MC
is a by-product of the periodic speed-up and slow-down of the
zonal flows.  Note that a complete study on the meridional flow changes 
across the cycle relies not only on the zonal Reynolds stresses
(Eq. \ref{eq.amf1}) but also on its meridional components 
\cite[][in preparation]{PMCG16}.

\section{Discussion and Conclusions}
\label{s.c}

Besides affecting the average profile of the differential rotation, 
the magnetic feedback generates TO in a periodic convective
dynamo model (model RC02).  We have demonstrated that the origin 
of these oscillations in our simulation is due to the magnetic torque
induced by the strong large-scale 
magnetic fields at the model's tachocline. The temporal evolution 
of the axial torques in different regions of our simulation domain 
suggests that the two branches of TO are directly driven by the magnetic 
tension (MT) at the base of the convection zone. This perturbation propagates
upwards up to the surface. 
The sign difference between the poles and the equator, as well as 
the latitudinal migration of the TO, is explained by the phase 
delay between the MT at higher and lower latitudes.

Despite the amplitude of the TO in our simulations being higher than the 
observed amplitude in the Sun, we notice morphological similarities
with the observations for both the TO and the variation of the MC.
Our results support the hypothesis that it is the magnetic field that modifies 
the meridional circulation during the solar cycle. 
This is in contrast to the idea that the meridional flow governs the solar cycle, 
as proposed by flux-transport dynamo models \citep[e.g.,][]{NMJM11}. In our 
results, the variations of the MC appear correlated 
with variations of the angular velocity, which, in turn, are driven by 
the deep dynamo-generated magnetic field. 

The distribution of the magnetic field below the photosphere is an 
open question and is relevant for the understanding of the solar dynamo. It
should be addressed with the correct understanding of the TO 
\citep[e.g.,][]{ACG13}. This work represents a step 
forward in that direction. 

\acknowledgments
We thank the anonymous referee for insightful comments that 
helped to improve the paper.
This work was partly funded by FAPEMIG grant APQ-01168/14 (GG),
FAPESP grant 2013/10559-5 (EMGDP), CNPq grant 306598/2009-4 (EMGDP),
NASA grants NNX09AJ85g and NNX14AB70G. 
PKS is supported by funding received from the
European Research Council under the European Union's Seventh Framework
Programme (FP7/2012/ERC Grant agreement no. 320375). The simulations
were performed in the NASA cluster Pleiades and the computing facilities
of the Laboratory of Astroinformatics (IAG/USP, NAT/Unicsul) supported by
a FAPESP (grant 2009/54006-4).

\bibliographystyle{apj}


\clearpage

\end{document}